\documentclass[preprint,showpacs,preprintnumbers,amsmath,amssymb,prb]{revtex4}
\usepackage{graphicx}
\usepackage{dcolumn}
\usepackage{bm}

\begin{document}

\title{Electronic properties and hyperfine fields of nickel-related
complexes in diamond}

\author{R. Larico$^{\rm (1)}$, J. F. Justo$^{\rm (2)}$,
W. V. M. Machado$^{\rm (1)}$, and L. V. C. Assali$^{\rm (1)}$}

\affiliation{$^{\rm (1)}$ Instituto de F\'{\i}sica,
Universidade de S\~ao Paulo,\\
CP 66318, CEP 05314-970, S\~ao Paulo, SP, Brazil \\
$^{\rm (2)}$ Escola Polit\'ecnica, Universidade de S\~ao Paulo,\\
CP 61548, CEP 05424-970, S\~ao Paulo, SP, Brazil}
%\date{\today}
%\maketitle

\begin{abstract}
We carried out a first principles investigation on the microscopic properties
of nickel-related defect centers in diamond. Several configurations, involving
substitutional and interstitial nickel impurities, have been considered either
in isolated configurations or forming complexes with other defects, such as
vacancies and boron and nitrogen dopants. The results, in terms of spin,
symmetry, and hyperfine fields, were compared with the available
experimental data on electrically active centers in synthetic diamond.
Several microscopic models, previously proposed to explain those
data, have been confirmed by this investigation, while
some models could be discarded. We also provided new insights on
the microscopic structure of several of those centers.
\end{abstract}

\maketitle

\section{introduction}
\label{sec1}

Diamond is a material which stands alone in nature, carrying a unique combination
of electronic, mechanical, thermal, and optical properties. Diamond is the hardest
known natural material, having a large bulk modulus, high thermal conductivity and
a large electronic band gap. Those properties make it a prototypical material
to a number of applications, ranging from drilling and cutting tools to electronic
devices to operate under extreme conditions \cite{isberg}. More recently,
new potential applications for doped diamond have been proposed, such as
superconducting materials \cite{lee} and quantum computing \cite{morton,kaxiras}.

There are two major methods with widespread use to grow macroscopic samples of
synthetic diamond. Chemical vapor deposition (CVD) methods produce high quality
diamond thin films grown over large areas. On the other hand, the high pressure-high
temperature (HPHT) processes produce bulk diamond at relatively high growth
rates and low costs. In those processes, samples are grown out of graphite, using
3d transition metal (TM) alloys (involving nickel, cobalt, and iron) as
solvent-catalysts. Nickel is the only impurity that has been unambiguously
identified in the resulting diamond. Such residual nickel impurities, either
isolated or forming complexes with other defects, can generate several electrically
and optically active centers \cite{yelisseyev}. Understanding the nature and
microscopic structure of those centers is crucial in developing diamond-related
technologies. Following the experimental identification of
those centers, several microscopic models have been proposed to explain such
data. However, there is still considerable controversy over a unified model
which could explain most of those active centers in diamond. Here,
we used first principles calculations to address this question.

Over the last decade, nickel-related impurities in diamond have been investigated
by several theoretical approaches \cite{gerstmann,main,larico,goss}. However, no
investigation has provided a comprehensive picture of most of the nickel-related
active centers identified in diamond so far. We used first principles total
energy calculations, based on the full-potential linearized augmented plane
wave methodology \cite{singh}, to investigate the structural
and electronic properties of those centers in terms of the spin, symmetry,
ground state multiplet, formation and transition energies, and hyperfine parameters.
We focused our investigation on centers involving isolated nickel, either in
interstitial or substitutional configurations, and complexes involving nickel and
vacancies or dopants (boron and nitrogen). This paper is organized as follow: in
section \ref{sec2} we discuss the available electron paramagnetic resonance (EPR)
experimental data of nickel-related defects in diamond. In section \ref{sec3},
we present the methodology used in this investigation. Sections \ref{sec4}
and \ref{sec5} present and discuss the results in the context of
experimental data.

\section{A survey on the experimental data}
\label{sec2}

\begin{figure}[ht]
\centering{
\includegraphics[width=160mm]{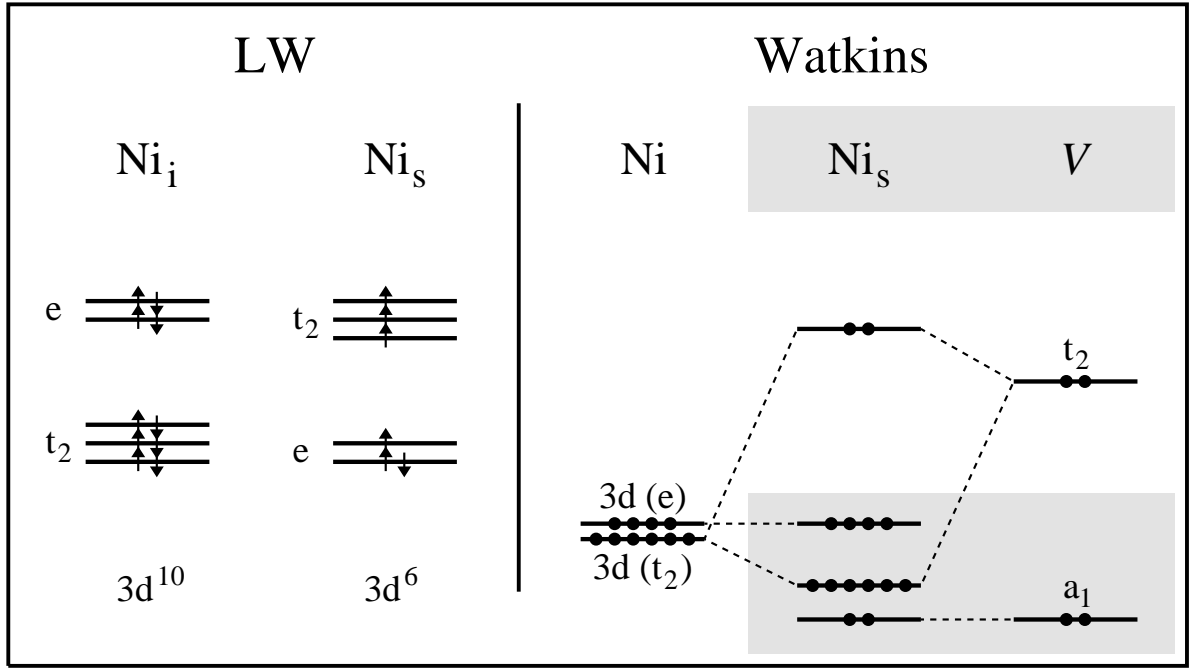}
\caption{Schematic representation of the gap states for
an isolated interstitial (Ni$_{\rm i}$) or substitutional (Ni$_{\rm s}$) nickel in
neutral charge state, according to LW \cite{lw} and Watkins (vacancy) \cite{watkins2}
models. The $\uparrow$ and $\downarrow$ arrows represent the spin up and
down, respectively. Gray regions represent the valence and conduction
host bands. For simplicity, the system is considered
in a tetrahedral symmetry, neglecting distortions.}
\label{fig1}
}
\end{figure}

Electron paramagnetic resonance and optical absorption measurements have been
used to identify a number of nickel-related active centers in diamond, and have
been recently reviewed \cite{yelisseyev}. The electronic properties of those
centers have been analyzed in terms of either the Ludwig-Woodbury (LW) \cite{lw}
or the vacancy \cite{watkins2} model. According to the LW model, when a
3d$^{\rm n}$4s$^2$ ion ($\rm 1\leq n < 9$) occupies an interstitial site in a
type-IV semiconductor, its 4s electrons are transferred to the 3d orbitals,
resulting in a 3d$^{\rm n+2}$ configuration. In the tetrahedral crystal field,
the 3d states are split into $e+t_2$ irreducible representations. The threefold
$t_2$ states lie lower in energy than the two-fold $e$ states. This level ordering
is the result of the octahedral crystal field, created by the next nearest
neighbors of the impurity, which is stronger than the tetrahedral crystal field
from the nearest neighbors. The same ion in a substitutional site would present a
3d$^{\rm n-2}$ configuration, since four electrons are needed to bind with the
four nearest neighboring host atoms. However, in this case the crystal field has
tetrahedral symmetry, driving the $e$ states to lie lower in energy than the
$t_2$ ones. Additionally, in the LW model, the gap levels are filled according
to the Hund's rule. The LW model is schematically presented in fig. \ref{fig1}.

For substitutional impurities, there is an alternative model, called the vacancy
model, proposed by Watkins for TM elements near the end of the 3d, 4d, and 5d
series \cite{watkins2}. This model proposed that the electronic structure of
the impurity resulted from a weak interaction between the impurity-related
d-$t_2$ states and the $t_2$ vacancy-related ones, as represented in
fig. \ref{fig1}. The vacancy-related states came from the dangling bonds on the
host atoms surrounding the vacant site into which the transition metal was
inserted. As a result, the impurity band gap states would have a vacancy-like
behavior. Although those two models were developed to describe the properties
of 3d-transition metal impurities in silicon, they have been extensively used to
explain the microscopic properties of those impurities in other semiconductors,
such as nickel-related impurity centers in diamond.

Nickel in diamond has been detected in a tetrahedral symmetry with a spin
S=3/2 by EPR \cite{isoya1} and optical measurements \cite{davies1}, and has been
labeled W8 center. The microscopic model suggested for this center, based on
either the LW or vacancy models, is an isolated substitutional nickel in the
negative charge state (Ni$_{\rm s}^-$) in a 3d$^7$ configuration \cite{isoya1}.

Two major active centers have been found in synthetic diamond, which have been
associated to interstitial nickel, labeled NIRIM-1 and NIRIM-2
centers \cite{isoya2}. The NIRIM-1 has been identified with a spin S=1/2 in a
trigonal symmetry at low temperatures (T$<$ 25 K), which switches to a
tetrahedral symmetry at higher temperatures. This center was discussed in the
context of the LW model, and interpreted as resulting from an isolated
interstitial nickel in the positive charge state (Ni$_{\rm i}^+$) \cite{isoya2}.
Since substitutional nickel in a positive charge state would give a spin S=5/2
according to the LW model, it was ruled out as a possible microscopic
configuration for the NIRIM-1 center. More recently, independent investigations
suggested that this center could, in fact, be formed by Ni$_{\rm s}^+$, giving
a spin S=1/2  \cite{larico,baker}, indicating that the vacancy model is more
suitable to describe this center.

The NIRIM-2 center has been identified with a spin S=1/2 with a strong trigonal
distortion \cite{isoya2,nazare1,watkins}. The microscopic structure of this
center is still the subject of controversy. It was initially associated to an
interstitial nickel with an impurity or vacancy nearby \cite{isoya2}. More
recently, this center has been proposed to be formed by a complex of nickel and
boron \cite{baker} or even by an isolated interstitial Ni \cite{larico}.

Post-growth annealing treatments introduce new active centers in the as-grown
samples, which have been labeled NE centers \cite{nadolinny}. It has been
suggested that those NE centers involve nickel, nitrogen and vacancies. The
NE4 center, that displays a D$_{3d}$ symmetry and a spin S=1/2, has been
tentatively associated to an interstitial Ni sitting in the middle position of
a divacancy. This vacancy-nickel-vacancy unit ($V$Ni$V$) would be aligned along
a $\langle 111 \rangle$ direction. This configuration has also been labeled as
NiC$_6$ in the literature, which represents, besides the impurity, the six
nearest neighboring carbon atoms. The LW model \cite{lw} was invoked to
describe this configuration, in which the Ni impurity would donate six of its
ten d-electrons to form bonds to the six neighboring carbon atoms. The
remaining four 3d electrons should occupy a triplet d-orbital, according
to Hund's rule. Since the center has a spin S=1/2, then this center should
be in the negative charge state ($V$Ni$V$)$^-$, associated to a $t^5_2$
electronic configuration. The NE4 center is the precursor to several other NE
centers, which are formed by replacing nearest neighboring carbon atoms with
nitrogen ones \cite{nadolinny}, in a NiC$_{\rm 6 -m}$N$_{\rm m}$
($1\leq {\rm m} \leq 6$) configuration. Recently, a center with rhombohedral
symmetry and spin S=1, has been observed in diamond \cite{iako}. A ($V$Ni$V$)
configuration was suggested for this center, similar to that of NE4, but here
in the neutral charge state. The NE1 center has a monoclinic symmetry and spin
S=1/2, and has been suggested to be formed by a ($V$Ni$V$) configuration plus
two nearest neighboring nitrogen atoms \cite{nadolinny2}. The NE8 center has a
monoclinic symmetry and spin S=1/2, and has been suggested to be formed by a
($V$Ni$V$) configuration plus four nearest neighboring nitrogen
atoms \cite{nadolinny2}.

EPR data has unambiguously shown that nickel can pair with boron and nitrogen
impurities in diamond, forming new active centers. A center, labeled NOL1,
has been identified with spin S=1 and trigonal symmetry \cite{nadolinny2}.
It has been suggested that this center is formed by an interstitial
$\rm Ni_i^{2+} (3d^8)$ impurity axially distorted by a boron ($\rm B_s^-$)
along a $\langle 111 \rangle$ direction, with an unspecified interatomic
distance between the impurities.  A more recent examination of the trigonal
boron-related NOL1 center suggested a different model, that would involve
substitutional nickel and boron, $\rm Ni_s^{+} B_s^0$, with the acceptor boron
in a next nearest neighboring site, with no covalent bonding between the
impurities \cite{baker}.

In samples with high concentrations of both nickel and nitrogen, other active
centers have been identified. In addition to the NE centers, the AB5 center,
with a spin S=1 and trigonal symmetry, has been identified \cite{neves}.
The microscopic model proposed for this center is a substitutional nickel
(Ni$_{\rm s}^{2-}$) with a nearby substitutional nitrogen atom (N$_{\rm s}^{+}$).
Table \ref{tab1} summarizes the properties of nickel-related EPR active
centers in diamond, as well as the respective proposed microscopic models.

Most of the microscopic models proposed in the previous paragraphs have been
built based on an ionic model \cite{zhao}, which has been proposed to describe
the 3d-transition metal-acceptor pairs $\rm TM_{\rm i}^+$-A$_{\rm s}^-$ in
silicon \cite{lw} (TM in a tetrahedral interstitial site in a positive charge
state plus an acceptor A in a negative charge state). According to that
model, the pair stable configuration corresponds to a classical system
consisting of a TM$_{\rm i}^+$ electrostatically bound to a nearest neighbor
A$_{\rm s}^-$ embedded in a dielectric medium \cite{assali7}. Since the
negatively charged acceptor has a closed shell, the electronic properties of
the pair can be directly related to the positive TM ion placed in a screened
Coulomb field.

\section{Methodology}
\label{sec3}

We used the all-electron spin-polarized full-potential linearized augmented
plane wave (FP-LAPW) method \cite{singh}, implemented in the WIEN2k
package \cite{blaha}. The calculations were performed within the framework of
the density functional theory, using the Perdew-Burke-Ernzerhof
exchange-correlation potential \cite{pbe}. All the calculations were performed
considering a 54-atom reference supercell. The methodology separates the
crystalline space in two distinct regions: the atomic and interstitial ones.
The electronic wave functions were expanded in terms of spherical harmonics in
the atomic regions and of plane waves in the interstitial ones. We chose all
atomic spheres with a radius of R =  0.64 \AA. Therefore, 2R was much smaller
than the crystalline interatomic distance of 1.54 \AA, such that atomic sphere
overlap was avoided even in the case of large atomic relaxations. We used a
2 $\times$ 2 $\times$ 2 grid to sample the irreducible Brillouin zone as well
as the $\Gamma$-point.

Convergence in the total energy was tested by varying the number of plane waves
describing the electronic wave-functions in the interstitial region, a 7.0/R
value provided converged results. Self-consistent interactions were performed
until total energy and the total charge in the atomic spheres changed by less
than $10^{-4}$ eV/atom and $10^{-5}$ electronic charges/atom between two
iterations, respectively. Additionally, the atomic positions were relaxed
until the forces were smaller than 0.02 eV/\AA. All those approximations and
convergence criteria have been shown to provide an accurate description of
several defect centers in semiconductors \cite{assali,ayres,bn}.

Formation and transition energies of all centers were computed using the
procedure discussed in Ref. \cite{assali}. The procedure required the total
energies of the respective defect center and the chemical potentials of carbon,
nitrogen and nickel. Those chemical potentials were computed using the total
energy of carbon in a diamond lattice, nitrogen in a N$_2$ molecule, and nickel
in a FCC lattice. In order to compute the hyperfine tensors, spin-orbit coupling
was included in a second-variational procedure. Additional information concerning
the calculation of hyperfine tensors is presented in Appendix.

\section{Results}
\label{sec4}

We considered most of the proposed microscopic models for the electrically
active centers described in table \ref{tab1}, as well as, other possible models
for those centers. Figure \ref{fig2} represents the diamond lattice in the
$(1\overline{1}0)$ plane, showing the possible sites in which the impurities
could be placed in the beginning of each simulation. All atomic positions
were later relaxed, according to convergence criteria discussed in the previous
section.

\subsection{Isolated Nickel}

Substitutional nickel in diamond was considered in several charge states,
with the results summarized in table \ref{tab2}. In the case of the neutral
charge state (Ni$_{\rm s}^0$), the center has no point-symmetry
(C$_1$), and presents a spin S=1. This configuration is only 0.1 eV
more stable than the center in a C$_{3v}$ symmetry (Ni$_{\rm s}^{0\ast}$).
Figure \ref{fig3} displays the induced energy eigenvalues in the
gap region for substitutional nickel impurity. The gap states of Ni$_{\rm s}$
are vacancy-like orbitals, consistent with the vacancy model \cite{watkins2}.

Table \ref{tab2} also presents the results for interstitial nickel.
In the positively charge state (Ni$_{\rm  i}^+$), the center was initially
simulated in a trigonal (C$_{3v}$) symmetry, in order to check if that
configuration could explain the properties of the NIRIM-2 center \cite{larico}.
In that symmetry, it presented an effective spin S=1/2 and an $^2{\rm E}$
multiplet ground state. By releasing the symmetry constraint, there was an
energy gain of about 0.2 eV, and the center distorted to a C$_{1h}$ symmetry.
This symmetry lowering was very small, corresponding to a distortion on the
nickel atom of only 0.06 \AA\ toward one of its second nearest neighbors,
breaking the trigonal symmetry. Figure \ref{fig4} compares the electronic
structure of the Ni$_{\rm i}^+$ center in both symmetries, showing that
although the symmetry lowering was small, there were strong effects in the
electronic structure of the center. These results show that the electronic
structure of interstitial nickel cannot be described by the LW model \cite{lw},
since the 3d-nickel related states remain resonant in the valence band,
leaving a hole in the perturbed valence band top. In trigonal symmetry, the
valence band top of the diamond crystal splits into an $a_1$ state, resonant
in the valence band, and an $e$ state, occupied by three electrons, inside the
gap. In the C$_{1h}$ symmetry, the $e$ gap states split further in an $a'$
and an $a''$ ones.

\subsection{Ni-vacancy complexes}

We initially considered an interstitial nickel paired with a nearest
neighboring vacancy (Ni$_{\rm i} V$), as suggested as a stable configuration
in several experiments \cite{isoya2}. However, this configuration was unstable
and the impurity moved toward the vacant site, forming a substitutional
nickel \cite{larico}. We additionally considered a substitutional nickel
paired with a nearest neighboring vacancy (Ni$_{\rm s} V$), but in the final
relaxed structure, the nickel remained in the middle position between two
vacancies ($V$Ni$V$). Figure \ref{fig5} presents the induced energy eigenvalues
of this last complex and table \ref{tab2} presents the respective properties.

The electronic structure of the $V$Ni$V$ complex cannot be described by the
LW model, as it has been recently suggested \cite{nadolinny}. Our results
indicate that the relevant electronic properties of this center should be
associated to divacancy-like orbitals, which appeared in the gap, while the
Ni-related orbitals remained resonant and inert inside the valence band.
On the other hand, the electronic structure is well described by the crystal
field theory, in which the electronic states can be interpreted as resulting
from an interaction between the divacancy states and those of the Ni atom.
The one-electron ground state structure of a diamond divacancy in D$_{3d}$
symmetry has the $a_{\rm 2u}^2 a_{\rm 1g}^2 e_{\rm u}^2 e_{\rm g}^0$
configuration. In that symmetry, the Ni 3d energy levels are split into
2$e_{\rm g}+a_{\rm 1g}$. When a Ni atom is placed in the middle position of a
divacancy, its $e_{\rm g}$ energy level interacts with the carbon dangling
bonds, leaving a fully occupied non-bonding $t_{\rm 2g}$-like
($e_{\rm g}+a_{\rm 1g}$) orbital inside the valence band. On the other hand,
the Ni $e_{\rm g}$ state interacts with the divacancy $e_{\rm g}$ gap level,
leaving the $e_{\rm g}$-bonding level in the valence band and the
$e_{\rm g}$-anti-bonding one unoccupied in the gap. The relevant electronic
properties of this center are related to the $e_{\rm u}$ divacancy-like
orbital, which remained in the gap bottom. In the positive and negative
charge states, the symmetry lowering (D$_{3d} \rightarrow {\rm C}_{2h}$) is
very weak and the splitting in the $e_{\rm u}$-related states is smaller
than 0.1 eV.

\subsection{Ni-B complexes}

\begin{figure}[ht]
\centering{
\includegraphics[width=160mm]{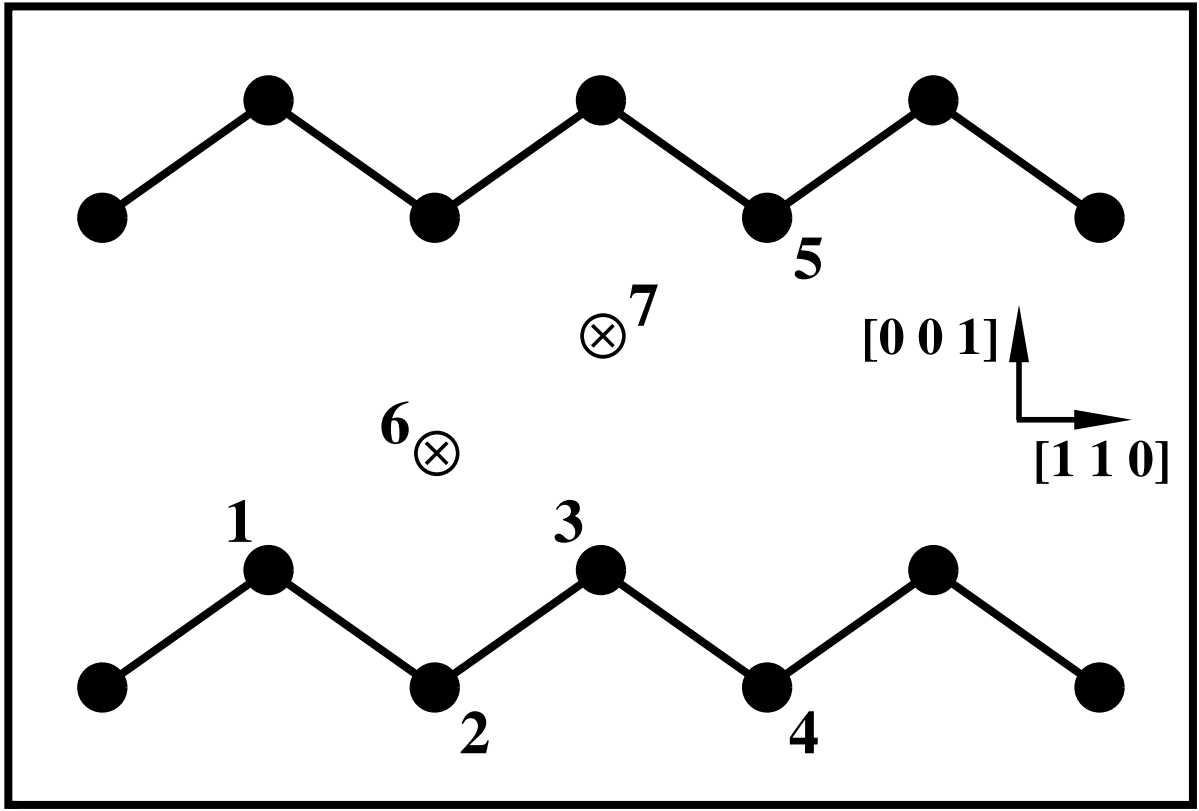}
\caption{Representation of the diamond lattice in the $(1\overline{1}0)$ plane.
Carbon atoms are represented by black circles. Labels, from 1 to 5, indicate the
crystal site positions where the impurities could be placed. The figure also
shows two tetrahedral interstitial sites (6 and 7), represented by
the $\bigotimes$ symbol, in the $[111]$ direction.}
\label{fig2}
}
\end{figure}

\begin{figure}[ht]
\centering{
\includegraphics[width=160mm]{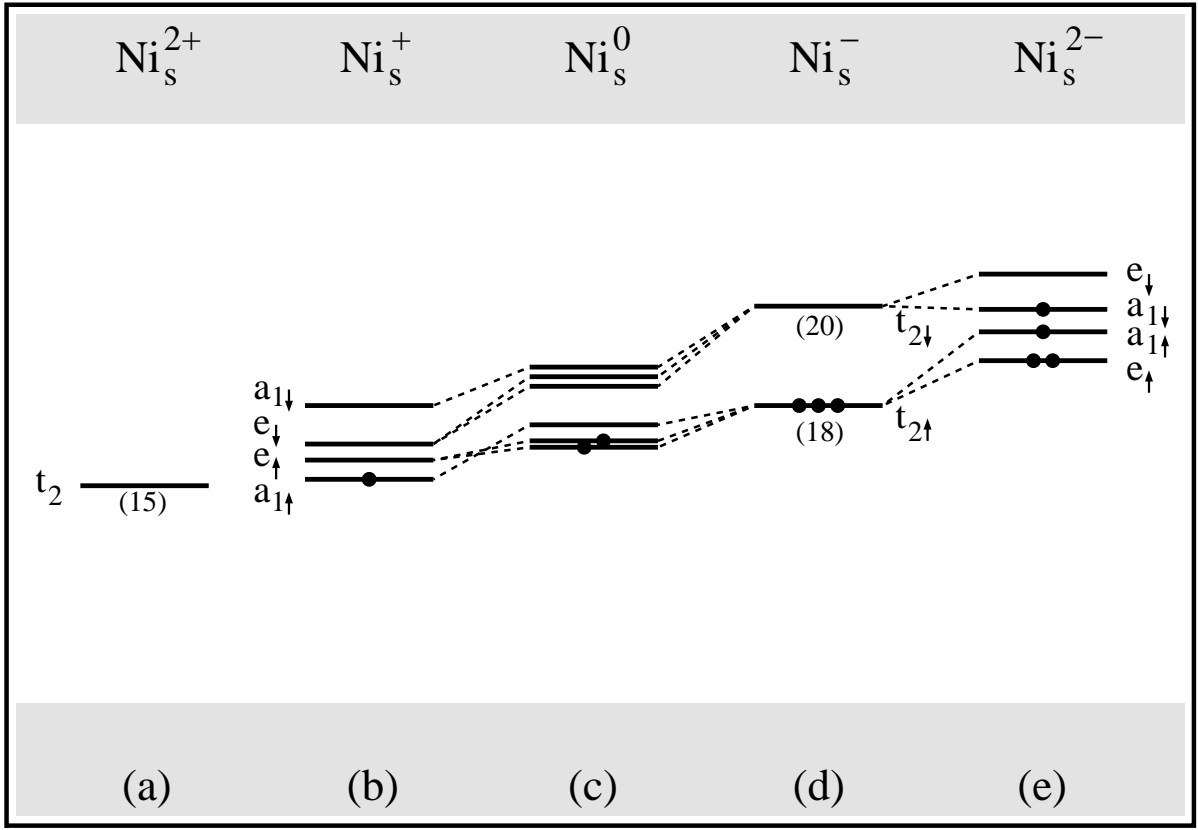}
\caption{The Kohn-Sham spin-polarized energy eigenvalues (around the
$\Gamma$ point) representing the 3$d$-related Ni levels in the gap
region for isolated substitutional nickel in different charge states:
(a) Ni${\rm _s^{2+}}$, (b) Ni${\rm _s^{+}}$, (c) Ni${\rm _s^{0}}$,
(d) Ni${\rm _s^{-}}$, and (e) Ni${\rm _s^{2-}}$. Levels with spin up and
down are represented by $\uparrow$ and $\downarrow$ arrows, respectively.
The occupation of the gap levels is given by the number of filled circles.
Numbers in parenthesis represent the $d$-character percentage of charge inside
the Ni atomic sphere.}
\label{fig3}
}
\end{figure}

\begin{figure}[ht]
\centering{
\includegraphics[width=120mm]{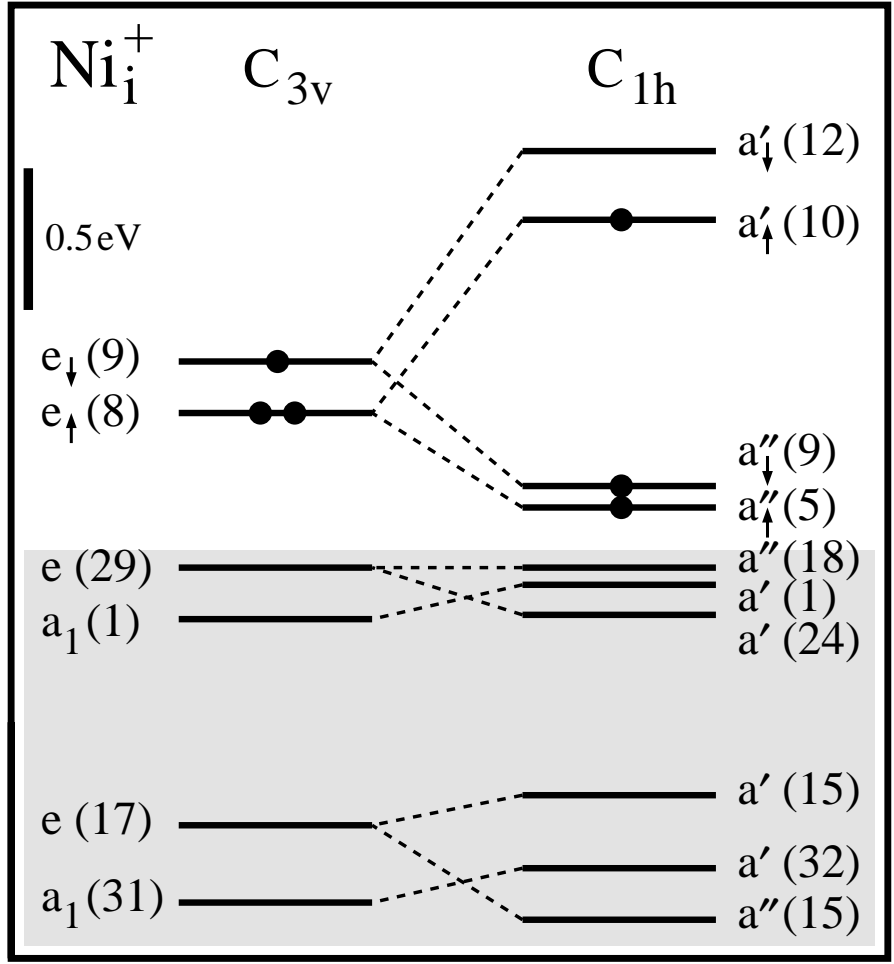}
\caption{The energy eigenvalues representing the 3$d$-related Ni levels for
isolated interstitial nickel in the positive charge state in C$_{3v}$ and C$_{1h}$
symmetries. Levels with spin up and down are represented by $\uparrow$ and
$\downarrow$ arrows, respectively. The occupation of the gap levels is given by the
number of filled circles. The numbers in parenthesis represent the $d$-character
percentage of charge inside the Ni atomic sphere.}
\label{fig4}
}
\end{figure}

\begin{figure}[ht]
\centering{
\includegraphics[width=160mm]{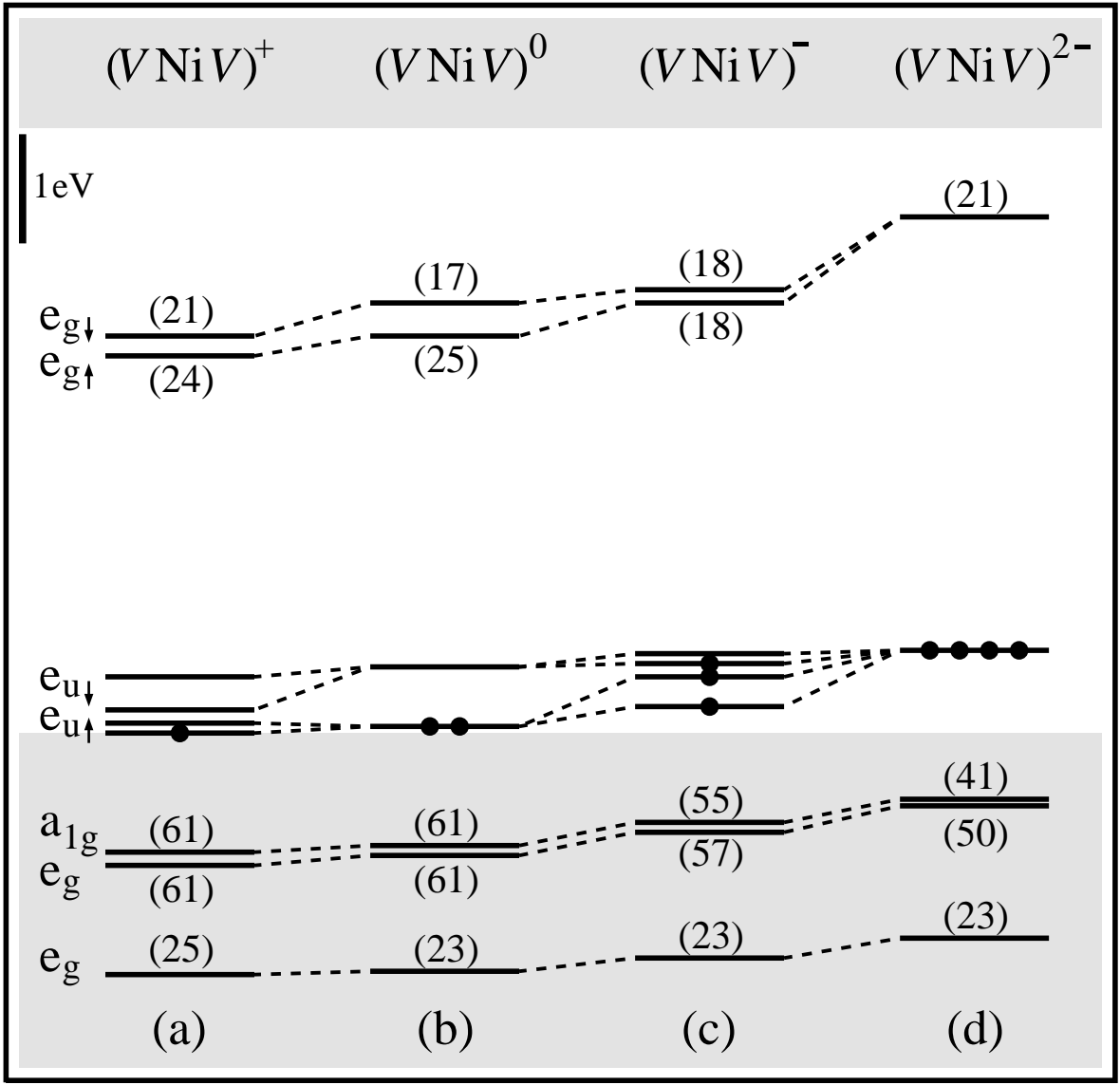}
\caption{The energy eigenvalues representing the 3$d$-related Ni levels in the gap
region for the Ni-divacancy complexes: (a) $(V{\rm Ni}V)^{+}$, (b) $(V{\rm Ni}V)^{0}$,
(c) $(V{\rm Ni}V)^{-}$, and (d) $(V{\rm Ni}V)^{2-}$ centers. The occupation of the gap
levels is given by the number of filled circles. The numbers in parenthesis represent
the $d$-character percentage of charge inside the nickel atomic sphere. Levels with spin
up and down are represented by $\uparrow$ and $\downarrow$ arrows, respectively.}
\label{fig5}
}
\end{figure}

\begin{figure}[ht]
\centering{
\includegraphics[width=130mm]{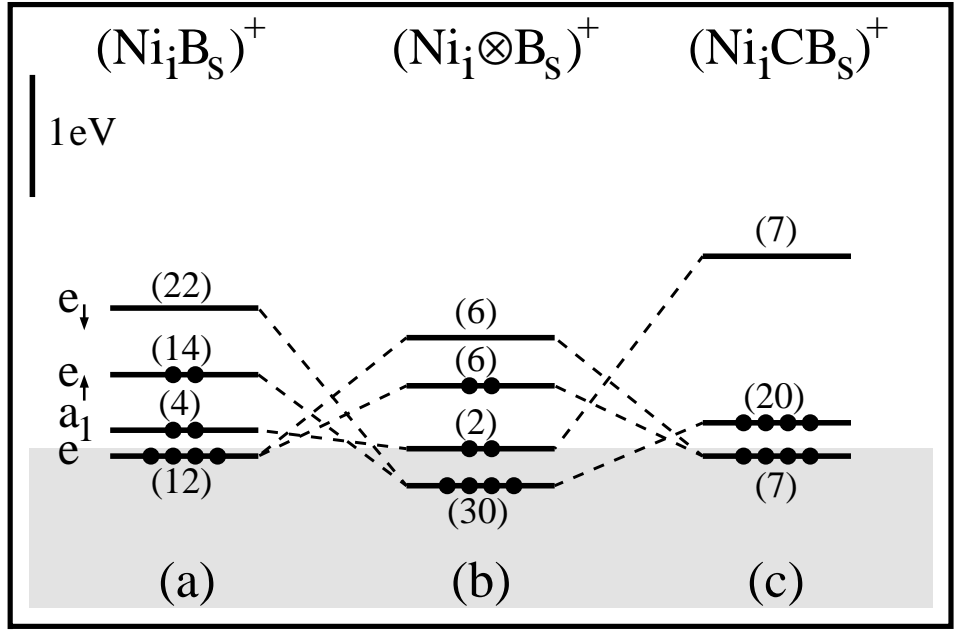}
\caption{The energy eigenvalues in the gap region for three configurations
involving interstitial nickel-substitutional boron complexes in the
positive charge state:
(a) $(\rm Ni_{i} B_{s})^+$, (b) $(\rm Ni_{i}\bigotimes B_{s})^+$, and
(c) $(\rm Ni_{i}C B_{s})^+$. The occupation of the levels is given by
the number of filled circles. The numbers in parenthesis represent the $d$-character
percentage of charge inside the nickel atomic sphere.}
\label{fig6}
}
\end{figure}

We now consider complexes involving
nickel and substitutional boron, which could potentially lead to a
trigonal symmetry, to be consistent with proposed models for Ni-B centers
presented in table \ref{tab1}. For interstitial nickel-substitutional boron
pairs, we considered three microscopic configurations, according to
figure \ref{fig2}: $\rm Ni_{i}B_{s}$ pair, with Ni and B respectively in sites
6 and 1; $\rm Ni_{i}CB_{s}$ pair, with Ni and B respectively in sites 6 and 4;
and $\rm Ni_{i} \! \otimes \! B_s$ with Ni and B respectively in sites 7 and 1.
Table \ref{tab3} presents the properties of those centers in several
charge states. The $\rm (Ni_{i}B_{s})^{0 \ast}$ complex, has a degenerate
configuration in C$_{3v}$ symmetry, coming from the partially occupied $e$
state, favoring a symmetry lowering to $\rm C_1$. The distance between the Ni
and B in the pairs is crucial for the final properties of those centers, as
evidenced by the electronic structure of those three centers, in the same
charge state, shown in figure \ref{fig6}. The major difference emerges on the
character of the highest occupied level in the center. While for the
$\rm Ni_{i}B_{s}$ and $\rm Ni_{i}CB_{s}$ pairs, this level has a localized Ni
3d-related character, for the $\rm Ni_{i} \! \otimes \! B_s$, this level is
essentially delocalized. For this last center, the distance between the
impurities is so large that the center can be well described by an ionic
model, in which the role of boron is only to accept an electron from the
nickel impurity. Therefore, the electronic structure of this complex can be
well described as an isolated interstitial nickel in 2+ charge state $\rm (Ni^{2+})$.

\begin{figure}[ht]
\centering{
\includegraphics[width=220mm,angle=90]{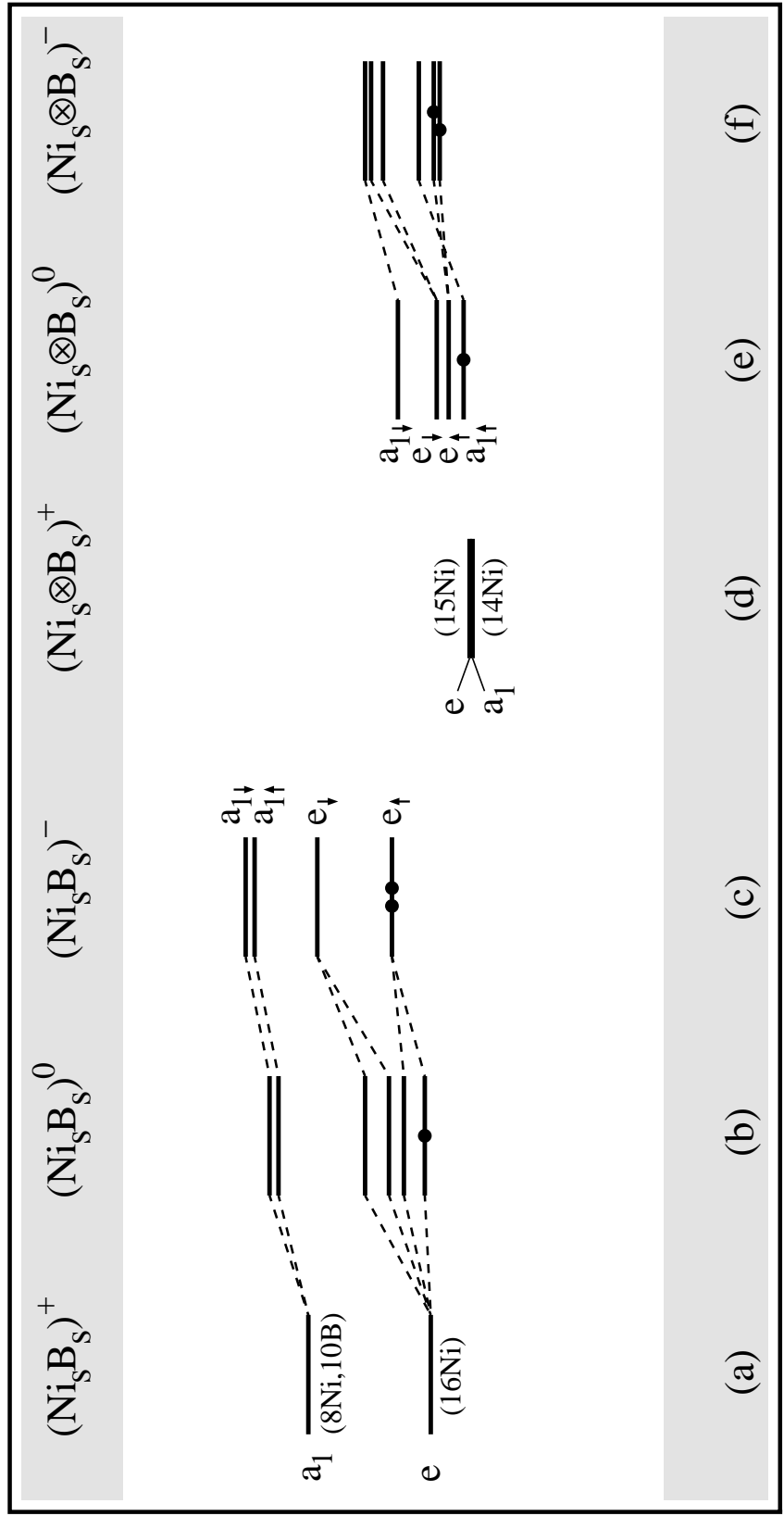}
\caption{The energy eigenvalues for the substitutional
nickel-substitutional boron complexes in two
configurations: (a,b,c) for the Ni$_{\rm s}$ B$_{\rm s}$ complex and (d,e,f)
for the Ni$_{\rm s} \bigotimes$ B$_{\rm s}$ complex. The occupation of
the gap levels is given by the number of filled circles. The numbers in parenthesis
represent the $d$-character ($p$-character) percentage of charge inside the nickel
(boron) atomic sphere. Levels with spin up and down are represented by $\uparrow$ and
$\downarrow$ arrows, respectively.}
\label{fig7}
}
\end{figure}

For substitutional nickel-substitutional boron, we considered two structural
configurations, according to figure \ref{fig2}: $\rm Ni_{s}B_{s}$ pair, with
Ni and B respectively in sites 3 and 4; $\rm Ni_{s}\! \otimes \! B_{s}$ pair,
with Ni and B respectively in sites 1 and 5. Figure \ref{fig7} presents the
energy eigenvalues of those pairs and table \ref{tab3} summarizes their
properties. In the $\rm Ni_{s} \! \otimes \! B_s$ centers, boron is far from
the nickel impurity, working as just an acceptor, such that the electronic
configuration resembles that of isolated substitutional Ni impurity,
shown in figure \ref{fig3}.
For the $\rm Ni_{s}B_{s}$ pair, boron plays a more important role, affecting
the electronic structure of the center, although the magnetic properties
of this center are associated with partially occupied energy levels
with prevailing nickel character.

\subsection{Ni-N complexes}

The nickel and nitrogen complexes in diamond are generally formed as result of
high-temperature thermal annealing, in which nitrogen impurities become highly
mobile and end up pairing with the less mobile nickel ones. We considered
centers with nickel in interstitial, substitutional, and divacancy sites
complexing with nitrogen. According to fig. \ref{fig2}, the $\rm Ni_{i}N_{s}$
center has Ni and N atoms respectively in sites 6 and 1 and the
$\rm Ni_{s}N_{s}$ center has Ni and N atoms respectively in sites 3 and 4.
Table \ref{tab4} presents the results for Ni-N pairs.

\begin{figure}[ht]
\centering{
\includegraphics[width=160mm]{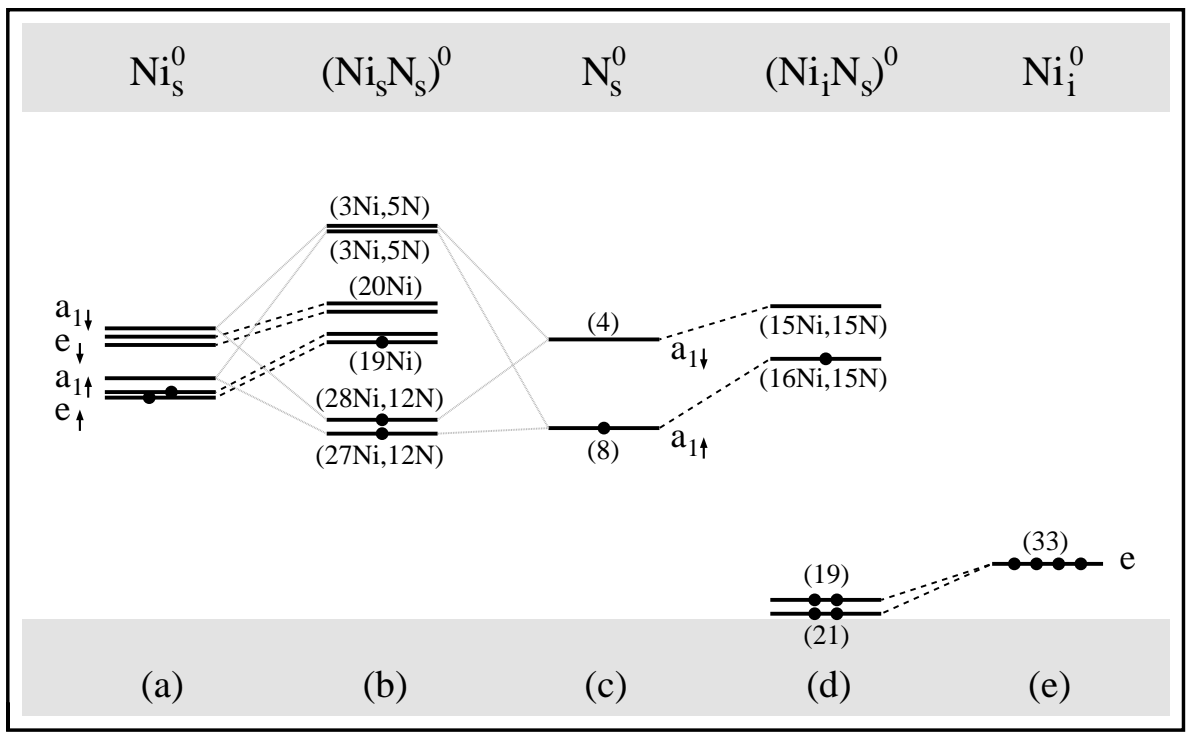}
\caption{The energy eigenvalues for
the (Ni$_{\rm i}$ N$_{\rm  s}$) and (Ni$_{\rm s}$ N$_{\rm s}$) complexes in
the neutral charge state. The figure shows that the electronic structure
of those centers results from hybridization between the $2p$
nitrogen with $3d$ nickel levels coming from its
precursors in isolated configurations.}
\label{fig8}
}
\end{figure}

Figure \ref{fig8} describes the electronic structure of Ni-N complexes as
resulting from an interaction of the energy eigenvalues of the impurities in
isolated configurations. For the $(\rm Ni_{s}N_{s})^0$ complex, there is a
hybridization between 3d-related gap states of Ni$_{\rm s}$ with those
2p-related ones of N$_{\rm s}$. This indicates a typical covalent interaction
between those two impurities \cite{assali0,justo0}. Despite this hybridization,
the highest occupied energy level in the $(\rm Ni_{s}N_{s})^0$ complex has an
$e$ representation, with a prevailing 3d character. On the other hand, for the
$(\rm Ni_{i}N_{s})^0$ complex, the electronic structure results from a
weaker interaction between the states of the isolated impurities, more
consistent with an ionic model \cite{zhao}. In this last case, the highest
occupied level has both 3d-related Ni and 2p-related N characters.

Table \ref{tab4} presents the results for complexes involving nickel and
nitrogen impurities in a divacancy site. We considered two possible
configurations, according to the proposed models for the NE1 and NE8 active
centers described in table \ref{tab1}. The $\rm N{\it V} Ni {\it V}N$
complex involves the precursor $\rm {\it V} Ni {\it V}$ plus two nitrogen
atoms in diametrically opposed positions, replacing two of the nickel six
nearest neighboring carbon atoms. The $\rm N_2{\it V} Ni {\it V}N_2$ complex
has four substitutional nitrogen atoms, replacing four of those nearest
neighboring carbon atoms. The electronic structure of those two centers shows
a strong covalent interaction between the divacancy-related orbitals and the
nitrogen-related ones, which is similar to what is observed for complexes
involving cobalt-nitrogen complexes in diamond \cite{larico11}. Nitrogen
incorporation into the precursor substantially alters the electronic structure
of that center. This shows that the current interpretation, in which nitrogen
atoms play a role of only donating electrons to the precursor, is not valid.

\section{Discussions and Conclusions}
\label{sec5}

The results are now discussed in the context of the experimentally identified
active centers observed in synthetic diamond. In a previous investigation, we
have shown that the microscopic configuration of a substitutional nickel in
the negative charge state (Ni$_{\rm s}^-$) was consistent with the properties
of the W8 center \cite{larico,larico10}, including symmetry, spin, and hyperfine
parameters. Additionally, we have shown that the previously proposed
microscopic models for the NIRIM centers, described in table \ref{tab1},
based on interstitial nickel impurities, were not consistent. For example,
the NIRIM-1 center could be better explained by an isolated substitutional
nickel in the positive charge state (Ni$_{\rm s}^+$) \cite{larico,baker}, although
this investigation finds a number of other configurations which are also
consistent with the NIRIM-1 symmetry and spin.

For the NIRIM-2 center, a direct comparison between theory and experiment was
more complicated. One of the proposed microscopic models for the NIRIM-2
center was an interstitial nickel with a nearby vacancy \cite{isoya2}.
Theoretical investigations showed that this configuration is
unstable \cite{larico}, such that the interstitial nickel migrates toward the
vacant site, becoming a substitutional impurity. This would be fully expected
considering defect energetics, since the formation energy of substitutional
nickel is considerably lower than that of an interstitial one. We have previously
suggested that isolated interstitial nickel in the positive charge state
(Ni$_{\rm  i}^+$) could explain some of the properties of the NIRIM-2
center \cite{larico}. This investigation shows that Ni$_{\rm  i}^+$ is
unstable in trigonal symmetry, lowering to a C$_1$ one. However, the energy
gain from this symmetry lowering is only 0.2 eV, and the final configuration
is not far from a trigonal symmetry. Recently, it has been proposed that
NIRIM-2 should involve interstitial nickel with a next nearest neighboring
boron atom \cite{baker}. The results for this proposed configuration,
(Ni$_{\rm i}$CB$_{\rm s})^0$, are fully consistent with the experimental data
for NIRIM-2 in terms of symmetry and spin. Another center, involving boron
and substitutional nickel, $(\rm Ni_s\!\otimes \! B_s)^0$, also provides
results consistent with experimental data of NIRIM-2. This last configuration
would be a strong candidate to explain the NIRIM-2 center since it involves
substitutional nickel, and formation energy is considerably smaller than that
for a pair involving interstitial nickel. A definite answer on the NIRIM-2
microscopic model could be achieved if future experiments could resolve the
Ni-hyperfine parameters, since according to table \ref{tab3}, those parameters
are considerably different for those two configurations.

Another center has been associated to nickel-boron pairs. The NOL1 center,
probably the same as the NIRIM-5 center, has been found in heavily boron-doped
diamond \cite{baker,nadolinny}. The center has trigonal symmetry and S=1.
By inspection of our results, the $(\rm Ni_i B_s)^+$ complex, suggested
by \cite{nadolinny} as the microscopic structure of this active center, is
fully consistent with the experimental data. Another complex involving
interstitial nickel, $(\rm Ni_i\! \otimes \! B_s)^+$, is also consistent with
experimental data. Complexes involving substitutional nickel could also
describe the properties of the NOL1 center. The $(\rm Ni_s B_s)^+$ complex,
suggested in Ref. \cite{baker}, is diamagnetic and cannot explain the NOL1
results. However, the same complex in a negative charge state,
$\rm (Ni_s B_s)^-$, is fully consistent with the experimental data. Although
this center has been only observed in heavily boron-doped diamond, our results
indicate that nickel in isolated configurations, Ni$_{\rm s}^{2-}$ or
Ni$_{\rm i}^{2+}$, are also consistent with the experimental data. In the
case of the Ni$_{\rm i}^{2+}$, the high concentration of boron would only be
required to place the Fermi level near the valence band top to access the
2+ charge state, and not necessarily participating into the complex. In
order to clarify this, EPR experiments should be performed to observe the
hyperfine parameters in nickel and boron nuclei.

For the NE4 centers, experiments \cite{nadolinny} have suggested a microscopic
structure given by a nickel impurity in a divacancy site in the negative
charge state ($V{\rm Ni} V$)$^-$. Our results for this configuration give a
trigonal symmetry and a spin S=1/2, both results consistent with the
experimental findings. The NE4* center\cite{iako}, from table \ref{tab1}, has been
suggested to be formed by $(V{\rm Ni}V)^0$. Our results corroborate
that suggestion, although they indicate a trigonal symmetry, while experiments
suggested a rhombohedral one. Moreover, our calculations found a hyperfine
parameter (A$_{\bot}$) of 60 MHz in the nearest neighboring carbon atoms,
very close to the experimental value of 79 MHz \cite{iako}.

The NE1 and NE8 centers have been suggested to be formed by nickel-nitrogen
complexes in a divacancy site \cite{nadolinny2}. Our results, in terms of
spin and symmetry, for the $\rm (N{\it V} Ni {\it V}N)^-$ and
$\rm (N_2{\it V} Ni {\it V}N_2)^+$ complexes are fully consistent with the
experimental data and the proposed microscopic configurations.
However, according to table \ref{tab4}, the
$\rm (N{\it V} Ni {\it V}N)^+$ and  $\rm (N_2{\it V} Ni {\it V}N_2)^-$  complexes
also provide results consistent with those data.
However, experiments \cite{nadolinny2} could not resolve the Ni-related hyperfine fields,
in order to compare with the values presented in table \ref{tab4}.
On the other hand, those experiments have identified hyperfine fields in the
nitrogen and the nearest neighboring carbon nuclei. For the NE1 center,
the experimental values for those fields are $\rm A_{\parallel} (N) = 59$,
$\rm  A_{\perp} (N) = 40$, $\rm A_{\parallel} (C) = 49$, and
$\rm  A_{\perp} (C) = 31$ MHz. For the $\rm (N{\it V} Ni {\it V}N)^-$ complex,
our results provide $\rm A_{\parallel} (N) = 42$ and
$\rm  A_{\perp} (N) = 17$ MHz and negligible values in the carbon nuclei.
For the $\rm (N{\it V} Ni {\it V}N)^+$ complex, our results provide
 $\rm A_{\parallel} (C) = 92$ and $\rm  A_{\perp} (C) = 40$ MHz and negligible
values in the nitrogen nuclei. Therefore, it was not possible
to make a final remark on the microscopic structure of the
NE1 center. For the $\rm (N_2{\it V} Ni {\it V}N_2)^+$ complex,
hyperfine fields in the nitrogen and carbon nuclei are fully consistent
with experimental values of the NE8 center.
Finally, the $(\rm Ni_s N_s)^-$ complex has been proposed as the microscopic
structure of the AB5 center \cite{neves}. From all the
complexes involving nickel and nitrogen considered here, that configuration
was the only one consistent with the experimental results of the AB5 center.

In summary, we have performed a theoretical investigation on nickel-related
complexes in diamond, in terms of electronic structure and hyperfine fields.
We have explored several microscopic configurations that
could explain the experimental data on EPR active centers in synthetic
diamond, confirming or discarding some of the previously proposed microscopic
models and suggesting new ones. These results provide a comprehensive picture
on Ni-related active centers in diamond using a single theoretical methodology.

\acknowledgments

The authors acknowledge partial support from Brazilian agency CNPq. The
calculations were performed using the computational facilities of the
CENAPAD and the LCCA-CCE (Universidade de S\~ao Paulo).

\section*{APPENDIX: CALCULATION OF THE HYPERFINE TENSORS}

The EPR data can provide important information related to electrically active
centers in semiconductors, such as symmetry, spin, gyromagnetic factor, and in
some stances the atomic composition of those centers. The hyperfine spectrum of
a center results from an interaction between nuclear magnetic moments
($\vec \mu_I$) and the moments of unpaired electrons. The crystalline field,
in which the impurity (or other defects) is immersed, is generally enough
strong to quench the respective orbital moment. However, it has been shown
that in the case of transition metal impurities, the orbital moment
is not fully quenched by the crystal field, generating, in some cases, a large
energy anisotropy. Most of the theoretical investigations have neglected this
anisotropic contribution, but it is very important for systems such as
those investigated here.

The hyperfine fields were computed using the implementation from the
WIEN2k package \cite{blaha} that uses a scalar-relativistic
approximation \cite{blugel}. According to that approach, the hyperfine magnetic
field ($\vec {\rm B}_{\rm hf}$) is computed considering three components: the
Fermi contact ($\vec {\rm  B}_{\rm c}$), the dipolar ($\vec {\rm B}_{\rm dip}$),
and the orbital ($\vec {\rm B}_{\rm orb}$) terms.
\begin{equation}
\vec {\rm B}_{\rm hf} =  \vec {\rm B}_{\rm c} +  \vec {\rm B}_{\rm dip} +
\vec {\rm  B}_{\rm orb},
\end{equation}

These three components are given in terms of the the angular ($\vec {\rm L}$) and
spin ($\vec {\rm S}$) electronic moments (in $\hbar$ unities) and the
Bohr magneton ($\beta_{\rm e} = {\rm e}\hbar / 2 m$):
\begin{eqnarray}
\vec{\rm B}_{\rm c} &=& \frac{8 \pi}{3} \beta_{\rm e}\, \vec{m}_{\rm av}, \\
\vec{\rm B}_{\rm dip} &=& - g_{\rm e} \beta_{\rm e} \, \langle \,\Phi \, \left|
\, \frac{S(r)}{r^3} \, \left[ \vec{\rm S}- 3 \left( \vec{\rm S}\cdot \vec{r}
\right) \, \frac{\vec{r}}{r^2} \right] \, \right| \, \Phi \, \rangle , \\
\vec{\rm B}_{\rm orb} &=& 2 \beta_{\rm e} \, \langle \, \Phi\, \left| \,
\frac{S(r)}{r^3}\, \vec{\rm L} \, \right| \, \Phi \, \rangle ,
\end{eqnarray}
where $\Phi$ is the relativistic large component of the wavefunction and
$S(r)$ is the reciprocal relativistic mass enhancement:
\begin{equation}
S(r) = \left[ 1 + \frac{\varepsilon - V(r)}{2mc^2} \right]^{-1},
\end{equation}
where $\varepsilon$ and $V(r)$ are respectively the kinetic energy and
the Coulomb potential.

$\vec{m}_{av}$ is the average nuclear magnetization,
\begin{eqnarray} \nonumber
\vec{m}_{\rm av} & = & \int \delta_{\rm T}({\vec {r}^{\, \prime}})
\vec{m}({\vec{r^{\, \prime}}})
{\rm d}{\vec  {r}^{\, \prime}} = \\
& = & \int \delta_{\rm T}({\vec{r}^{\, \prime}})  \langle \,
\Phi\, \left| \,\vec{\sigma} \,\delta({\vec {r}-\vec {r}^{\, \prime}})\,
\right| \, \Phi \, \rangle
\, {\rm d}{\vec{r}^{\, \prime}}
\end{eqnarray}
where $\delta_{\rm T}({\vec{r}^{\, \prime}})$ is given in terms of the Thomas
radius (${\rm r}_{\rm T} = {Z {\rm e}^2}/{mc^2}$):
\begin{equation}
\delta_{\rm T}({\vec r^{\, \prime}}) = \frac{1}{4\pi r^2}
\frac{\rm r_{T}}{\left[2r(1+\varepsilon/2 m c^2)+{\rm r_{T}}\right]}
\end{equation}
and $\vec{\sigma}$ are the Pauli matrices.

The splitting in energy resulting from the interaction between
the hyperfine magnetic fields ($\vec{\rm B}_{\rm hf}$) and
${\vec \mu}_{\rm I}$ is described by:
\begin{equation}
 {\rm E} = - {\vec \mu}_{\rm I} \cdot  {\vec{\rm B}}_{\rm hf}.
\end{equation}

This splitting in energy may be described in terms of a spin Hamiltonian (H).
The eigenvalues of this Hamiltonian provide information on the separation
between absorption lines in the magnetic spectra.
\begin{equation}
{\rm H}\, = \vec{\rm J} \cdot \tensor {\rm A} \cdot  \vec{\rm I} \,=\,
 (\vec{\rm L} + \vec{\rm S})\, \cdot \tensor {\rm A} \cdot  \vec{\rm I}
\end{equation}
where $\vec{\rm I}$ is the nuclear spin and $\tensor {\rm  A}$
is a ($3 \times 3$) hyperfine interaction tensor.

The hyperfine interaction tensor has the following components $\rm A_{ij}$:
\begin{eqnarray}
\rm {\rm A}_{ij} & =&  a^{\rm c}_{\rm ij} \delta_{\rm ij}
+ a^{\rm dip}_{\rm ij} +  a^{\rm orb}_{\rm ij}, \
\quad \mbox{with} \\ \nonumber
& & \sum_{\rm i} a^{\rm dip}_{\rm ii} =  0 \quad \mbox{and} \quad
\sum_{\rm i} a^{\rm orb}_{\rm ii} \neq 0.
\end{eqnarray}

In an experiment, when the direction of the external static magnetic field
($\hat {\rm n}  = \sin \theta \cos \varphi\, \hat {\imath} +\sin \theta \sin
\varphi\, \hat {\jmath} +  \cos \theta\,  \hat {k} $) is varied with respect
to the sample axis, the relevant quantity is the projection of the hyperfine
interaction tensor in that direction:
\begin{eqnarray} \nonumber
& & {\rm {A}} {\rm (\theta,\varphi)} \, =  \,
{\rm {\hat n}\, \cdot \, \tensor {\rm A} \, \cdot \, {\hat n}} = \\ \nonumber
& = & {\rm A_{11} \sin^2 \theta \cos^2 \varphi + (A_{12} + A_{21}) \sin \theta
\cos \varphi \sin  \varphi \, +} \\ \nonumber
&\, + & {\rm A_{22} \sin^2 \theta \sin^2 \varphi +
(A_{23} + A_{32}) \cos \theta \sin \theta \sin \varphi \, +}\\
&\, + & {\rm A_{33} \cos^2 \theta +
(A_{13} + A_{31}) \cos \theta \sin \theta \cos \varphi .}
\end{eqnarray}

By choosing a convenient set of six  directions, {\it i.e.} six sets of
$(\theta,\varphi)$, the values of ${\rm {A (\theta,\varphi)}}$ in those
directions allow to build the hyperfine interaction tensor. It can be later
diagonalized to obtain the three principal values, also called hyperfine
parameters ($\rm A_1,\ A_2,\ and\ A_3$), and their respective eigenvectors.

The hyperfine tensor $\tensor {\rm A}$ is given in terms of:
\begin{equation}
 \tensor {\rm A} =  {\it a}^{\rm c}{\openone} + \tensor {\rm B} + \tensor {\rm C},
\end{equation}
where $\openone$ is the unitary tensor, and $a^{\rm c}$ is the contact term,
$\tensor {\rm B}$ is a traceless anisotropic tensor related to
the dipolar interaction, and $\tensor {\rm C}$ is an anisotropic tensor
related to the orbital interaction.

If the angular magnetic moment is quenched, the isotropic part of the hyperfine
tensor is exactly the Fermi contact interaction and the anisotropic part is
the dipolar interaction. However, if the angular moment is not quenched, there
should be a contribution from this interaction to the hyperfine tensor changing
both the dipolar and contact terms. In this investigation, we observed that
the hyperfine orbital field is generally relevant and cannot be neglected.
This was result of the spin-orbit coupling in the 3d localized orbitals,
which are deformed due to the crystalline field.

\pagebreak

\pagebreak

\begin{table}[ht]
\caption{Experimental data on the electrically active centers of Ni-related
defects in diamond. The table presents the symmetry, spin, and proposed microscopic
model. X represents an unknown specie (a vacancy or impurity), and $V$
represents a vacancy. A list of additional relevant active centers in diamond
can be found in Reference \cite{isoya5}.}
\label{tab1}
\begin{center}
\begin{tabular}{llcc}
\hline \hline
Label     &   \ \ \ \ \ Sym.             &\ \ \ \  S \ \ \ \ & Model  \\
\hline
W8        & tetrahedral          &  3/2  & Ni$_{\rm s}^-$ $^{(a)}$\\
NIRIM-1   & trigonal (T$<25$K)   &  1/2  &
Ni$_{\rm i}^+$ $^{(b)}$, Ni$_{\rm s}^+$ $^{(c),(d)}$ \\
NIRIM-2  \ \  & trigonal         &  1/2  &
Ni$_{\rm i}^+$-X $^{(b)}$, $\rm Ni_i^+C B_s^-$ $^{(d)}$ \\
NE4       & trigonal             &  1/2  & ($V$Ni$V$)$^-$ $^{(e)}$ \\
NE4*      & rhombohedral          &   1   & ($V$Ni$V$)$^0$  $^{(f)}$ \\
NE1       & monoclinic           &  1/2  &
($ {\rm N}V{\rm Ni}V{\rm N}$)$^-$ $^{(g)}$ \\
NE8       & monoclinic           &  1/2  &
($ {\rm N}_2V{\rm Ni}V{\rm N}_2 $)$^+$ $^{(g)}$ \\
NOL1      & trigonal             &  1    &
$\rm Ni_s^{+} B_s^0$ $^{(c)}$, $\rm Ni_i^{2+} B_s^-$  $^{(g)}$  \\
AB5       & trigonal             &  1    &
Ni$_{\rm s}^{2-}$N$_{\rm s}^{\rm +}$ $^{(h)}$ \\
\hline \hline
\end{tabular}
\end{center}
$^{(a)}$ Reference \cite{isoya1}, $^{(b)}$ Reference \cite{isoya2},
$^{(c)}$ Reference  \cite{larico}, $^{(d)}$ Reference \cite{baker}, \\
$^{(e)}$ Reference \cite{nadolinny}, $^{(f)}$ Reference \cite{iako},
$^{(g)}$ Reference \cite{nadolinny2},
$^{(h)}$ Reference \cite{neves}.
\end{table}
\pagebreak

\begin{table}[ht]
\caption{Results for isolated Ni and Ni-divacancy complexes in diamond:
symmetry, spin (S), multiplet ground state ($^{\rm 2S+1}\Gamma$), formation energies
(E$_{\rm F}$), and transition energies (E$_{\rm t}$ with relation to the valence
band top $\varepsilon_{\rm v}$). Here $\epsilon_{\rm F}$ is the Fermi energy.
The table also presents the calculated hyperfine parameters (A$_{\rm i}$, i =
1, 2, 3) in the $^{\rm 61}$Ni nucleus. Energies and hyperfine parameters are
given in eV and MHz, respectively.}
\label{tab2}
\vspace*{0.2cm}
\begin{center}
\begin{tabular}{lcccccccc}
\hline \hline
Center  & {~Sym.~} & {~~~S~~~} & {~~~$^{\rm 2S+1}\Gamma$~~~} &
{~~~E$_{\rm F}$~~~}   & E$_{\rm t}$ & A$_{1}$  & A$_{2}$ &  A$_{3}$ \\
\hline
Ni$_{\rm s}^{2+}$& ~~~T$_{\rm d}$~~~  & 0  & $^1{\rm A}_1$  &
~~3.9 + 2{\large {$\epsilon$}}$_{\rm F}$~~  & ~2.0 $(2+/+)$~& --  & -- & -- \\
Ni$_{\rm s}^+$ & C$_{3v}$  &  1/2 & $^2{\rm A}_1$ &
5.9 + {{$\epsilon$}}$_{\rm F}$ & 2.6 $(+/0)$  &  ~~123~~ & ~~-36~~ & ~~-36~~ \\
Ni$_{\rm s}^{0\,*}$ & C$_{3v}$&  1 & $^3{\rm E}$ & 8.6  & & 52 & 9 & 9 \\
Ni$_{\rm s}^0 $ & C$_{1}$&  1 & $^3{\rm A}$   & 8.5  &  & 45 & 18 & 4 \\
Ni$_{\rm s}^-$ & T$_{\rm d}$ &  3/2 & $^4{\rm A}_2$ &
 11.5 - {{$\epsilon$}}$_{\rm F}$ & 3.0 $(0/-)$ & 18 & 18 & 18\\
Ni$_{\rm s}^{2-}$ & C$_{3v}$ &  1 & $^3{\rm A}$ &
 15.5 - {{2$\epsilon$}}$_{\rm F}$ & 4.0 $(-/2-)$ & -99 & 21  & 21 \\[0.3cm]
Ni$_{\rm i}^{2+}$ &C$_{3v}$ & 1 & $^3{\rm A}$   &
15.5 + {2{$\epsilon$}}$_{\rm F}$ &  $0.6\,(2+/+)$  & 32  & 2 & 2  \\
Ni$_{\rm i}^{+\,*}$ & C$_{3v}$ & 1/2 & $^2{\rm E}$   &
16.3 + {{$\epsilon$}}$_{\rm F}$ &  & 29 & 15 & 15\\
Ni$_{\rm i}^+$ & C$_{1h}$ & 1/2 & $^2{\rm A}$   &
16.1 + {{$\epsilon$}}$_{\rm F}$ &  1.1 $(+/0)$  & 66 & 19 & 17 \\
Ni$_{\rm i}^0$ & T$_{\rm d}$  & 0  & $^1{\rm A}_1$ & 17.2 & & -- &-- & --\\[0.3cm]
($V{\rm Ni} V$)$^+$ & C$_{2h}$ & 1/2 & $^2$A &
5.2 + {{$\epsilon$}}$_{\rm F}$ & $0.2\,(+/0$)  &  51 & 17 & 14\\
($V{\rm Ni} V$)$^0$ &  D$_{3d}$ &  1  & $^3$A$_{2u}$ &
5.4  &  & 6 & 33 & 33\\
($V{\rm Ni} V$)$^-$ &  C$_{2h}$ & 1/2 & $^2$A &
6.2 - {{$\epsilon$}}$_{\rm F}$ & 0.8 $(0/-)$ & 18 & -52 & -22 \\
($V{\rm Ni} V$)$^{2-}$ &  D$_{3d}$ & 0 & $^1$A &
7.3 - 2{{$\epsilon$}}$_{\rm F}$ & 1.1 $(-/2-)$ &
-- & -- & --\\
\hline \hline
\end{tabular}
\end{center}
\end{table}
\vspace{1cm}
\pagebreak

\begin{table}[ht]
\caption{Results for nickel-boron complexes in diamond: symmetry, spin (S),
multiplet ground state ($^{\rm 2S+1}\Gamma$), formation (E$_{\rm F}$) and
transition energies (E$_{\rm t}$ with relation to $\varepsilon_{\rm v}$).
The table also presents the calculated hyperfine parameters (A$_{\rm i}$)
in the $^{\rm 61}$Ni nucleus.}
% and  $^{\rm 11}$B nuclei.}
\label{tab3}
\vspace*{0.2cm}
\begin{center}
\begin{tabular}{lcccccccc}
\hline \hline
Center  & {Sym.} & {~~S~~} & {~~~$^{\rm 2S+1}\Gamma$~~~} &
{~~~E$_{\rm F}$~~~} & E$_{\rm t}$ &  ~~A$_{1}$~~  & ~~A$_{2}$~~ &  ~~A$_{3}$~~ \\
%A$_{\parallel}$(Ni) & A$_{\perp}$(Ni) & A$_{\parallel}$(B) & A$_{\perp}$(B) \\
\hline
(Ni$_{\rm i}$B$_{\rm s})^+$ & C$_{3v}$ & 1 &  ~~~$^3{\rm A}$~~~ &
~$14.8 + {\epsilon}_{\rm F}$~&~1.1 $(+/0)$~& 7 & 35 & 35 \\% 0 & -2 \\
(Ni$_{\rm i}$B$_{\rm s})^{0\, *}$ & C$_{3v}$ & 1/2 & $^2{\rm E}$ &
16.1 &  &  76 & -62 & -62 \\ % & \\
(Ni$_{\rm i}$B$_{\rm s})^{0}$ & C$_{1}$ & 1/2 & $^2{\rm A}$ &
15.9 &  &  40 & -56 & -27 \\ % & \\
(Ni$_{\rm i}$B$_{\rm s})^-$ \ \ & C$_{3v}$ & 0 &  $^1{\rm A}_1$ &
~~$17.3 - {\epsilon}_{\rm F}$~~   &  ~1.4 $(0/-)$~  & -- & -- & -- \\[0.3cm] % & -- \\[0.3cm]
(Ni$_{\rm i}$CB$_{\rm s})^+$  \ & C$_{3v}$ & 0 &  $^1{\rm A}_1$ &
$15.4+{\large {\epsilon}}_{\rm F}$ &  ~1.8 $(+/0)$~    & -- & -- & -- \\ %& -- \\
(Ni$_{\rm i}$CB$_{\rm s})^0$  \ & C$_{3v}$ & 1/2 &  $^2{\rm A}_1$ &
17.2   &    & 21  & -14  & -14 \\ %16 & -7 \\
(Ni$_{\rm i}$CB$_{\rm s})^-$ & C$_{3v}$ & 0 & $^1{\rm A_1}$ &
$ 18.2 - {\epsilon}_{\rm F}$ & 1.4 $(+/-)$ & -- & -- & -- \\[0.3cm] %& -- \\[0.3cm]
$(\rm Ni_i\! \otimes \! B_s)^+$& C$_{3v}$ & 1 & $^3$A &
15.9 $+ {\epsilon}_{\rm F}$   &   0.5 $(+/0)$ &  7 & -17 & -17 \\ %&  &  \\
$(\rm Ni_i\! \otimes \! B_s)^0$ & C$_1$ & 1/2 &  $^2$A &
16.4 & & -5 & 50 & 46 \\ % & & \\
$(\rm Ni_i\! \otimes \! B_s)^-$ & C$_{3v}$ & 0 &  $^1{\rm A}_1$&
$17.9 - {\epsilon}_{\rm F}$ & 1.5 $(0/-)$  & -- & -- & -- \\[0.3cm] %  & & \\[0.3cm]
(Ni$_{\rm s}$B$_{\rm s})^+$ & C$_{3v}$ & 0 &  $^1{\rm A}_1$ &
$3.5 + {\epsilon}_{\rm F}$   &  2.6 $(+/0)$ & -- & -- & -- \\ %& -- & -- \\
(Ni$_{\rm s}$B$_{\rm s})^0$ & C$_{1}$ & 1/2 & $^2{\rm A}$ &
6.1  &  & -99 & 55 & -32  \\ % & & \\
(Ni$_{\rm s}$B$_{\rm s})^-$ \ \ & C$_{3v}$ & 1 &  $^3{\rm A}_1$ &
$9.0 - {\epsilon}_{\rm F}$   &  2.9 $(0/-)$  & 85 & 1 & 1 \\[0.3cm] %  & 7 & 4 \\[0.3cm]
$(\rm Ni_s\! \otimes \! B_s)^+$& C$_{3v}$ & 0 &  $^1{\rm A}_1$ &
$4.7 + {\epsilon}_{\rm F}$   &  2.1 $(+/0)$ & -- & -- & -- \\ % & -- & -- \\
$(\rm Ni_s\! \otimes \! B_s)^0$ & C$_{3v}$ & 1/2 & $^2{\rm A}_1$ &
6.8  & & -120 & 48 & 48 \\  %& & \\
$(\rm Ni_s\! \otimes \! B_s)^-$ & C$_{1}$ & 1 &  $^3{\rm A}$ &
$9.2 - {\epsilon}_{\rm F}$ & 2.4 $(0/-)$  & 42  & 18 & 8 \\ % & & \\
\hline \hline
\end{tabular}
\end{center}
\end{table}
\vspace{1cm}
\pagebreak

\begin{table}[ht]
\caption{Results for nickel-nitrogen complexes in diamond: symmetry, spin (S),
multiplet ground state ($^{\rm 2S+1}\Gamma$), formation  (E$_{\rm F}$) and transition
energies (E$_{\rm t}$ with relation to $\varepsilon_{\rm v}$). The table also
presents the calculated hyperfine parameters (A$_{\rm i}$)
in the $^{\rm 61}$Ni nucleus.}
\label{tab4}
\vspace*{0.2cm}
\begin{center}
\begin{tabular}{lcccccccc}
\hline \hline
Center  & {~Sym.~} & {~~~S~~~} & {~$^{\rm 2S+1}\Gamma$~} &
{~~~E$_{\rm F}$~~~}  & E$_{\rm t}$ &   ~~A$_{1}$~~  & ~~A$_{2}$~~ &
~~A$_{3}$~~ \\
%A$_{\parallel}$(Ni) & A$_{\perp}$(Ni) & A$_{\parallel}$(N) & A$_{\perp}$(N) \\
\hline
(Ni$_{\rm i}$N$_{\rm s})^+$ \ \ & ~~C$_{3v}$~~ & 0 &  ~~$^1{\rm A}_1$~~ &
15.8 + {{$\epsilon$}}$_{\rm F}$  &  3.2 $(+/0)$ & -- & -- & -- \\
(Ni$_{\rm i}$N$_{\rm s})^0$ & C$_{3v}$ & 1/2 & $^2{\rm A}_1$ &
19.0  &  & 82 & 14 & 14 \\
(Ni$_{\rm i}$N$_{\rm s})^-$ \ \ & C$_{3v}$ & 0 &  $^1{\rm A}_1$ &
~22.5 - {{$\epsilon$}}$_{\rm F}$~ & ~3.5 $(0/-)$~ & -- & -- & -- \\[0.3cm]
(Ni$_{\rm s}$N$_{\rm s})^+$ \ \ & C$_{3v}$ & 0 &  $^1{\rm A}_1$ &
6.0 + {{$\epsilon$}}$_{\rm F}$   &  3.1 $(+/0)$  & -- & -- & -- \\
(Ni$_{\rm s}$N$_{\rm s})^0$ & C$_{1}$ & 1/2 & $^2{\rm A}$ &
 9.1 &  & -110 & 70 & -30 \\
(Ni$_{\rm s}$N$_{\rm s})^-$ \ \ & C$_{3v}$ & 1 &  $^3{\rm A}_1$ &
12.6 - {{$\epsilon$}}$_{\rm F}$  & 3.5 $(0/-)$  & -58 & 2 & 2 \\ [0.3cm]
$\rm (N{\it V} Ni {\it V}N)^+$  & C$_{2h}$ & 1/2 &  $^2{\rm A}$ &
3.2 + {{$\epsilon$}}$_{\rm F}$   &  1.3 $(+/0)$  & 28  & -20  & -24 \\
$\rm (N{\it V} Ni {\it V}N)^0$  & C$_{2h}$ & 0 &  $^1{\rm A}$ &
4.5    &   & -- & -- & -- \\
$\rm (N{\it V} Ni {\it V}N)^-$  & C$_{2h}$ & 1/2 &  $^2{\rm A}$ &
7.5 - {{$\epsilon$}}$_{\rm F}$   &  3.0 $(0/-)$  &  177 & 58 & 43 \\ [0.3cm]
$\rm (N_2{\it V} Ni {\it V}N_2)^+$ \ \ & C$_{2h}$ & 1/2 &  $^2{\rm A}$ &
2.7 + {{$\epsilon$}}$_{\rm F}$          &  3.7 $(+/0)$  & -167 & -18 & -15 \\
$\rm (N_2{\it V} Ni {\it V}N_2)^0$ \ \ & C$_{2h}$ & 0 &  $^1{\rm A}$ &
6.4    &   & -- & -- & -- \\
$\rm (N_2{\it V} Ni {\it V}N_2)^-$ \ \ & C$_{2h}$ & 1/2 &  $^2{\rm A}$ &
10.7 - {{$\epsilon$}}$_{\rm F}$          &  4.3 $(0/-)$  & 1  & -5  & -4 \\
\hline \hline
\end{tabular}
\end{center}
\end{table}
\pagebreak

\pagebreak

\end{document}